\documentclass[conference]{IEEEtran}
\IEEEoverridecommandlockouts
\usepackage{cite}
\usepackage{amsmath,amssymb,amsfonts, mathtools}
\usepackage{algorithmic}
\usepackage{graphicx}
\usepackage{textcomp}
\usepackage{xcolor}
\def\BibTeX{{\rm B\kern-.05em{\sc i\kern-.025em b}\kern-.08em
    T\kern-.1667em\lower.7ex\hbox{E}\kern-.125emX}}

\usepackage[acronym]{glossaries}
\usepackage{siunitx}
\usepackage{booktabs}
\usepackage{multirow}
\usepackage{url}
\usepackage[flushmargin,hang]{footmisc}

\newcommand{\rev}[1]{\textcolor{black}{#1}} 
\newcommand{\revs}[1]{\textcolor{black}{#1}}

\begin{document}

\title{NanoHydra: Energy-Efficient \\ Time-Series Classification at the Edge
\thanks{
$*$ Equal contributions. \\ $\mathparagraph$ Original methodology proponent, currently with WiTricity.}

}

\author{
\IEEEauthorblockN{Cristian Cioflan\textsuperscript{$*$}\textsuperscript{$\dagger$}, Jos\'e Fonseca\textsuperscript{$*$}\textsuperscript{$\ddagger$}\textsuperscript{$\mathparagraph$}, Xiaying Wang\textsuperscript{$\dagger$}, Luca Benini\textsuperscript{$\dagger$}\textsuperscript{$\mathsection$}
}
\IEEEauthorblockA{
$\dagger$ \textit{Integrated Systems Laboratory, ETH Zurich, Switzerland} \\
$\ddagger$ \textit{Dept. of Information Technology and Electrical Engineering, ETH Zurich, Switzerland} \\
$\mathsection$ \textit{Dept. of Electrical, Electronic and Information Engineering, University of Bologna, Italy} \\
\{cioflanc, xiaywang, lbenini\}@iis.ee.ethz.ch, jcastro@student.ethz.ch}
}

\maketitle

\begin{abstract}

Time series classification (TSC) on extreme edge devices represents a stepping stone towards intelligent sensor nodes that preserve user privacy and offer real-time predictions.
Resource-constrained devices require efficient TinyML algorithms that prolong the device lifetime of battery-operated devices without compromising the classification accuracy.
We introduce NanoHydra, a TinyML TSC methodology relying on lightweight binary random convolutional kernels to extract meaningful features from data streams.
We demonstrate our system on the ultra-low-power GAP9 microcontroller, exploiting its eight-core cluster for the parallel execution of computationally intensive tasks.
We achieve a classification accuracy of up to 94.47\% on ECG5000 dataset, comparable with state-of-the-art works.
Our efficient NanoHydra requires only \qty{0.33}{\milli \second} to accurately classify a 1-second long ECG signal.
With a modest energy consumption of \qty{7.69}{\mu \joule} per inference, 18$\times$ more efficient than the state-of-the-art, NanoHydra is suitable for smart wearable devices, enabling a device lifetime of over four years.

\end{abstract}


% Generics
\newacronym{lpwan}{LPWAN}{Low-Power Wide Area Network}
\newacronym{lora}{LoRa}{Long Range}
\newacronym{lorawan}{LoRaWAN}{Long Range Wide Area Network}
\newacronym{nbiot}{NB-IoT}{Narrow Band Internet-of-Things}
\newacronym[plural=WANs, firstplural={Wide Area Networks (WANs)}]{wan}{WAN}{Wide Area Network}
\newacronym[plural=WSNs, firstplural={Wireless Sensor Networks (WSNs)}]{wsn}{WSN}{Wireless Sensor Network}
\newacronym{simd}{SIMD}{Single Instruction Multiple Data}
\newacronym{os}{OS}{Operating System}
\newacronym{ble}{BLE}{Bluetooth Low-Energy}
\newacronym{wifi}{Wi-FI}{Wireless Fidelity}
\newacronym[plural=DVS, firstplural={Dynamic Vision Sensors (DVS)}]{dvs}{DVS}{Dynamic Vision Sensor}
\newacronym{ptz}{PTZ}{Pan-Tilt Unit}

% Computer Architecture
\newacronym[plural=FLLs,firstplural=Frequency Locked Loops (FLLs)]{fll}{FLL}{Frequency Locked Loop}
\newacronym{dram}{DRAM}{Dynamic Random Access Memory}
\newacronym{fpu}{FPU}{Floating Point Unit}
\newacronym{dma}{DMA}{Direct Memory Access}
\newacronym[plural=LUTs, firstplural={Lookup Tables (LUTs)}]{lut}{LUT}{Lookup Table}
\newacronym[plural=FPGAs, firstplural={Field Programmable Gate Arrays (FPGAs)}]{fpga}{FPGA}{Field Programmable Gate Array}
\newacronym{dsp}{DSP}{Digital Signal Processing}
\newacronym{mcu}{MCU}{Microcontroller Unit}
\newacronym{spi}{SPI}{Serial Peripheral Interface}
\newacronym{cpi}{CPI}{Camera Parallel Interface}
\newacronym{fifo}{FIFO}{First-In First-Out Queue}
\newacronym{uart}{UART}{Universal Asynchronous Receiver-Transmitter}
\newacronym{raw}{RAW}{Read-After-Write}
\newacronym[plural=ISAs, firstplural={Instruction Set Architectures (ISAs)}]{isa}{ISA}{Instruction Set Architecture}

% Quantization 
\newacronym{ste}{STE}{Straight-Through-Estimator}

\newacronym[plural=PTUs, firstplural={Pan-Tilt Units}]{ptu}{PTU}{Pan-Tilt Unit}
\newacronym{mdf}{MDF}{Medium-density fibreboard}
\newacronym{cvat}{CVAT}{Computer Vision Annotation Tool}
\newacronym{coco}{COCO}{Common Objects in Context}
\newacronym{soa}{SoA}{State of the Art}
\newacronym{sf}{SF}{Sensor Fusion}

% Deep Learning generics
\newacronym{dl}{DL}{Deep Learning}
\newacronym{bn}{BN}{Batch Normalization}
\newacronym{FGSM}{FBK}{Fast Gradient Sign Method}
\newacronym{lr}{LR}{Learning Rate}
\newacronym{sgd}{SGD}{Stochastic Gradient Descent}
\newacronym{gd}{GD}{Gradient Descent}

\newacronym{sta}{STA}{Static Timing Analysis}

\newacronym[plural=GPIOs, firstplural={General Purpose Inupt Outputs (GPIOs)}]{gpio}{GPIO}{General Purpose Input Output}
\newacronym[plural=LDOs, firstplural={Low Dropout Regulators (LDOs)}]{ldo}{LDO}{Low Dropout Regulator}

\newacronym{inq}{INQ}{Incremental Network Quantization}

\newacronym{CV}{CV}{Computer Vision}
\newacronym{EoT}{EoT}{Expectation over Transformation}
\newacronym{RPN}{RPN}{Region Proposal Network}
\newacronym{TV}{TV}{Total Variation}
\newacronym{NPS}{NPS}{Non-Printability Score}
\newacronym{STN}{STN}{Spatial Transformer Network}
\newacronym{MTCNN}{MTCNN}{Multi-Task Convolutional Neural Network}
\newacronym{YOLO}{YOLO}{You Only Look Once}
\newacronym{SSD}{SSD}{Single Shot Detector}
\newacronym{SOTA}{SOTA}{State of the Art}
\newacronym{NMS}{NMS}{Non-Maximum Suppression}
\newacronym{ic}{IC}{Integrated Circuit}
\newacronym{rf}{RF}{Radio Frequency}
\newacronym{tcxo}{TCXO}{Temperature Controlled Crystal Oscillator}
\newacronym{jtag}{JTAG}{Joint Test Action Group industry standard}
\newacronym{swd}{SWD}{Serial Wire Debug}
\newacronym{sdio}{SDIO}{Serial Data Input Output}
% \newacronym{ldo}{LDO}{Linear Dropout Regulator}

\newacronym[plural=PCBs, firstplural={Printed Circuit Boards (PCB)}]{pcb}{PCB}{Printed Circuit Board}
\newacronym[plural=ASICs, firstplural={Application Specific Integrated Circuits}]{asic}{ASIC}{Application Specific Integrated Circuit}

\newacronym[plural=SCMs, firstplural={Standard Cell Memories (SCMs)}]{scm}{SCM}{Standard Cell Memory}
\newacronym{ann}{ANN}{Artificial Neural Networks}
\newacronym{ml}{ML}{Machine Learning}
\newacronym{ai}{AI}{Artificial Intelligence}
\newacronym{iot}{IoT}{Internet of Things}
\newacronym{fft}{FFT}{Fast Fourier Transform}
\newacronym[plural=OCUs, firstplural={Output Channel Compute Units (OCUs)}]{ocu}{OCU}{Output Channel Compute Unit}
\newacronym{alu}{ALU}{Arithmetic Logic Unit}
\newacronym{mac}{MAC}{Multiply-Accumulate}
\newacronym{soc}{SoC}{System-on-Chip}
\newacronym{tcdm}{TCDM}{Tightly-Coupled Data Memory}
\newacronym{pulp}{PULP}{Parallel Ultra Low Power}
\newacronym{ulp}{ULP}{Ultra-Low-Power}
\newacronym{fc}{FC}{Fabric Controller}
\newacronym{ne16}{NE16}{Neural Engine 16-channels}

\newacronym{PGD}{PGD}{Projected Gradient Descend}
\newacronym{CW}{CW}{Carlini-Wagner}
\newacronym{OD}{OD}{Object Detection}

\newacronym{rrf}{RRF}{RADAR Repetition Frequency}
\newacronym{nlp}{NLP}{Natural Language Processing}
\newacronym{qam}{QAM}{Quadrature Amplitude Modulation}
\newacronym{rri}{RRI}{RADAR Repetition Interval}
\newacronym{radar}{RADAR}{Radio Detection and Ranging}
\newacronym{loocv}{LOOCV}{Leave-one-out cross validation}

\newacronym{nas}{NAS}{Neural Architecture Search}
\newacronym{bsp}{BSP}{Board Support Package}
\newacronym{ttn}{TTN}{The Things Network}
\newacronym{wip}{WIP}{Work in Progress}
\newacronym{json}{JSON}{JavaScript Object Notation}
\newacronym{qat}{QAT}{Quantization-Aware Training}

% Metrics
\newacronym{cls}{CLS}{Classification Error}
\newacronym{loc}{LOC}{Localization Error}
\newacronym{bkgd}{BKGD}{Background Error}
\newacronym{roc}{ROC}{Receiver Operating Characteristic}
\newacronym{frr}{FRR}{False Rejection Rate}
\newacronym{eer}{EER}{Equal Error Rate}
\newacronym{snr}{SNR}{Signal-to-Noise Ratio}
\newacronym{flop}{FLOP}{Floating-Point Operation}
\newacronym{fp}{FP}{Floating-Point}
\newacronym{fps}{FPS}{Frames Per Second}

% Datasets
\newacronym{gsc}{GSC}{Google Speech Commands}
\newacronym{mswc}{MSWC}{Multilingual Spoken Words Corpus}
\newacronym{demand}{DEMAND}{Diverse Environments Multichannel Acoustic Noise Database}
\newacronym{kinem}{KINEM}{Keywords In Noisy Environments and Microphones}

% Topologies
\newacronym[plural=BNNs, firstplural={Binary Neural Networks (BNNs)}]{bnn}{BNN}{Binary Neural Network}
\newacronym[plural=NNs, firstplural={Neural Networks}]{nn}{NN}{Neural Network (NNs)}
\newacronym[plural=SNNs, firstplural={Spiking Neural Networks (SNNs)}]{snn}{SNN}{Spiking Neural Network}
\newacronym[plural=DNNs, firstplural={Deep Neural Networks (DNNs)}]{dnn}{DNN}{Deep Neural Network}
\newacronym[plural=TCNs,firstplural=Temporal Convolutional Networks]{tcn}{TCN}{Temporal Convolutional Network}
\newacronym[plural=CNNs,firstplural=Convolutional Neural Networks (CNNs)]{cnn}{CNN}{Convolutional Neural Network}
\newacronym[plural=TNNs,firstplural=Ternarized Neural Networks]{tnn}{TNN}{Ternarized Neural Network}
\newacronym{ds-cnn}{DS-CNN}{Depthwise Separable Convolutional Neural Network}
\newacronym{rnn}{RNN}{Recurrent Neural Network}
% \newacronym{cnn}{CNN}{Convolutional Neural Network}
\newacronym{gcn}{GCN}{Graph Convolutional Network}
\newacronym{mhsa}{MHSA}{Multi-Head Self Attention}
\newacronym{crnn}{CRNN}{Convolutional Recurrent Neural Network}
\newacronym{clca}{CLCA}{Convolutional Linear Cross-Attention}
% \newacronym{mhsa}{MHSA}{Multi-Head Self-attention}
\newacronym{resnet}{ResNet}{Residual Network}

% Audio methods
\newacronym{bf}{BF}{Beamforming}
\newacronym{anc}{ANC}{Active Noise Cancellation}
\newacronym{agc}{AGC}{Automatic Gain Control}
\newacronym{se}{SE}{Speech Enhancement}
\newacronym{mct}{MCT}{Multi-Condition Training}
\newacronym{mcta}{MCTA}{Multi-Condition Training \& Adaptation}
\newacronym{pcen}{PCEN}{Per-Channel Energy Normalization}
\newacronym{mfcc}{MFCC}{Mel-Frequency Cepstral Coefficient}
\newacronym{asr}{ASR}{Automated Speech Recognition}
\newacronym{kws}{KWS}{Keyword Spotting}
\newacronym{odl}{ODL}{On-Device Learning}
\newacronym{odcl}{ODCL}{On-Device Continual Learning}

% Paper specific
\newacronym{hydra}{HYDRA}{HYbrid Dictionary-Rocket Architecture}
\newacronym{rck}{RCK}{Random Convolutional Kernel}
\newacronym{rocket}{ROCKET}{RandOm Convolutional KErnel Transform}
\newacronym{ecg}{ECG}{electrocardiogram}
\newacronym{adc}{ADC}{Analog-to-Digital Converters}
\newacronym{tsc}{TSC}{Time Series Classification}
\newacronym{svm}{SVM}{Support Vector Machine}
\newacronym{knn}{k-NN}{k-Nearest Neighbors}
\newacronym{cpu}{CPU}{Central Processing Unit}
\newacronym{ulpm}{ULPM}{Ultra-Low-Power Mode}
\newacronym{lpm}{LPM}{Low-Power Mode}
\newacronym{hpm}{HPM}{High-Performance Mode}
\newacronym{ulpws}{ULPWS}{Ultra-Low-Power Wearable System}

\newcommand{\Ndl}{N_{\text{dil}}}
\newcommand{\Ndf}{N_{\text{diff}}}
\newcommand{\Vbat}{V_{\text{bat}}}
\newcommand{\Vcore}{V_{\text{core}}}
\newcommand{\Iavg}{I_{\text{avg}}}
\newcommand{\Pavg}{P_{\text{avg}}}
\newcommand{\Pinf}{P_{\text{inf}}}
\newcommand{\Psl}{P_{\text{sleep}}}
\newcommand{\Padc}{P_{\text{ADC}}}
\newcommand{\Pactive}{P_{\text{active}}}
\newcommand{\Pidle}{P_{\text{idle}}}
\newcommand{\ti}{\Delta t_{\text{inf}}}
\newcommand{\ta}{\Delta t_{\text{acq}}}
\newcommand{\mah}{\qty{}{\milli\ampere\hour}}
\newcommand{\Bcap}{B_{\text{cap}}}
\newcommand{\Lh}{L_{\text{hours}}}
\newcommand{\Ly}{L_{\text{years}}}
\newcommand{\ndl}{n_{\text{dil}}}
\newcommand{\ndf}{n_{\text{diff}}}

\section{Introduction}
\label{sec:introduction}

\gls{tsc} is the \gls{ml} task of predicting categorical class labels for windows of data streams, which are time-indexed and ordered.
\gls{tsc} algorithms are employed in a plethora of real-world applications such as \gls{ecg} analysis to assess heart conditions~\cite{ingolfsson2021ecgtcn}, human activity classification~\cite{raj2023improved}, brain-machine interfaces~\cite{mei2024train}, or speech recognition~\cite{cioflan2024ondevice}.

\rev{A highly desirable} characteristic of \gls{tsc} systems is their ability to process and classify sensor data directly on (extreme) edge devices -- intelligent sensor nodes.
This enables the preservation of user privacy and \rev{curtails} the energy consumption and the latency \rev{overheads} associated with client-server communication \rev{of raw data}.
In such smart edge computing systems, the classification accuracy does not represent the only metric quantifying the performance of a \gls{ml} algorithm. 
The reduced computational budget, limited battery lifetime, and limited onboard memory~\cite{lin2023tiny} must be addressed individually to achieve a real-time classification solution suitable for resource-constrained~\glspl{mcu}, as defined by the TinyML paradigm.

Advances in the computational capabilities of edge devices and deployment tools~\cite{scherer2024deeploy} encouraged the development of complex~\gls{dnn} models solving~\gls{tsc} tasks, such as \glspl{cnn}~\cite{raj2023improved,wang2024adafsnet, ay2022study,cioflan2024ondevice, mei2024train}, \glspl{tcn}~\cite{ingolfsson2021ecgtcn}, \glspl{rnn}~\cite{xiao2021rnts}, or transformer models~\cite{shi2021self, scherer2024wip}.
However, in the context of battery-powered edge nodes performing real-time classification, lightweight approaches are instead required to prolong the lifetime of the devices.

\gls{rocket} models~\cite{dempster2020rocket, dempster2021minirocket, giordano2023minirocket} have been recently proposed as alternatives to traditional \glspl{dnn} solving \gls{tsc} tasks. 
Such topologies rely on~\glspl{rck} to extract information from input data streams, which are further processed by linear classifiers to produce an accurate prediction.
The low computational requirements of \gls{rocket} solutions \rev{contribute to reducing} both the training and inference time, making them suitable for extreme edge devices such as the NRF52810 \gls{mcu}, as demonstrated by~\cite{giordano2023minirocket}.
Dempster et al.~\cite{dempster2023hydra} introduced~\gls{hydra}, which employs dictionary methods to form groups of patterns that approximate the time series, the classification step representing the selection of the most frequent pattern observed in the input data. 
Unlike traditional dictionary methods, the grouped patterns in \gls{hydra} are produced by the \gls{rocket} random convolutional kernels.
~\gls{hydra}~\cite{dempster2023hydra} outperforms \gls{rocket}-based models on multiple UCR~\cite{hoanganh2018ucr} datasets, but its extreme edge feasibility has not yet been explored.

This work introduces NanoHydra\footnote{https://github.com/pulp-platform/nanohydra}, a TinyML approach suitable for extreme edge devices performing accurate time series classification.
Integerized \glspl{rck} are sampled from the \{-1,1\} set, simultaneously reducing the memory and computational costs.
Furthermore, our proposed algorithm replaces costly operations such as square-root computations with arithmetic shifts, thus reducing the computational complexity without compromising the accuracy.
We achieve accuracy levels up to 94.47\% on the ECG5000~\cite{hoanganh2018ucr} dataset, comparable with state-of-the-art models~\cite{lo2024timeseriesclassificationrandom} evaluated on multi-core CPUs with GBs of RAM.
We deploy NanoHydra on the ultra-low-power GAP9 \gls{mcu}, exploiting its eight-core cluster by parallelizing the convolutions, as well as the \gls{simd} extensions for the vectorization of the integer operations.
We achieve real-time on-device inference within \qty{0.33}{\milli \second}, with a \rev{low end-to-end energy budget} of \qty{7.69}{\mu \joule}, demonstrating time series classification on intelligent sensor nodes.
We estimate the device lifetime of TinyML platforms running NanoHydra, showing that continuous operation between three months and over four years can be achieved.

\begin{figure*}[t]

\caption{\textbf{NanoHydra overview.} The input time series are first convolved with $\Ndl$ and $\Ndf$ versions of itself, producing the time series projections of length $L_{x}$. The projections are convolved with $H \times K$ random convolutional kernels. The result generates a feature vector of length $L_{F}$, produced after hard (i.e., in red) and soft (i.e., in gray) counting the convolutional output. The feature vector is convolved with the classifier's trainable weights, and the classification output is determined by selecting the index of the largest activation.}
% \vspace{0.3cm}
\centering
\includegraphics[width=15cm]{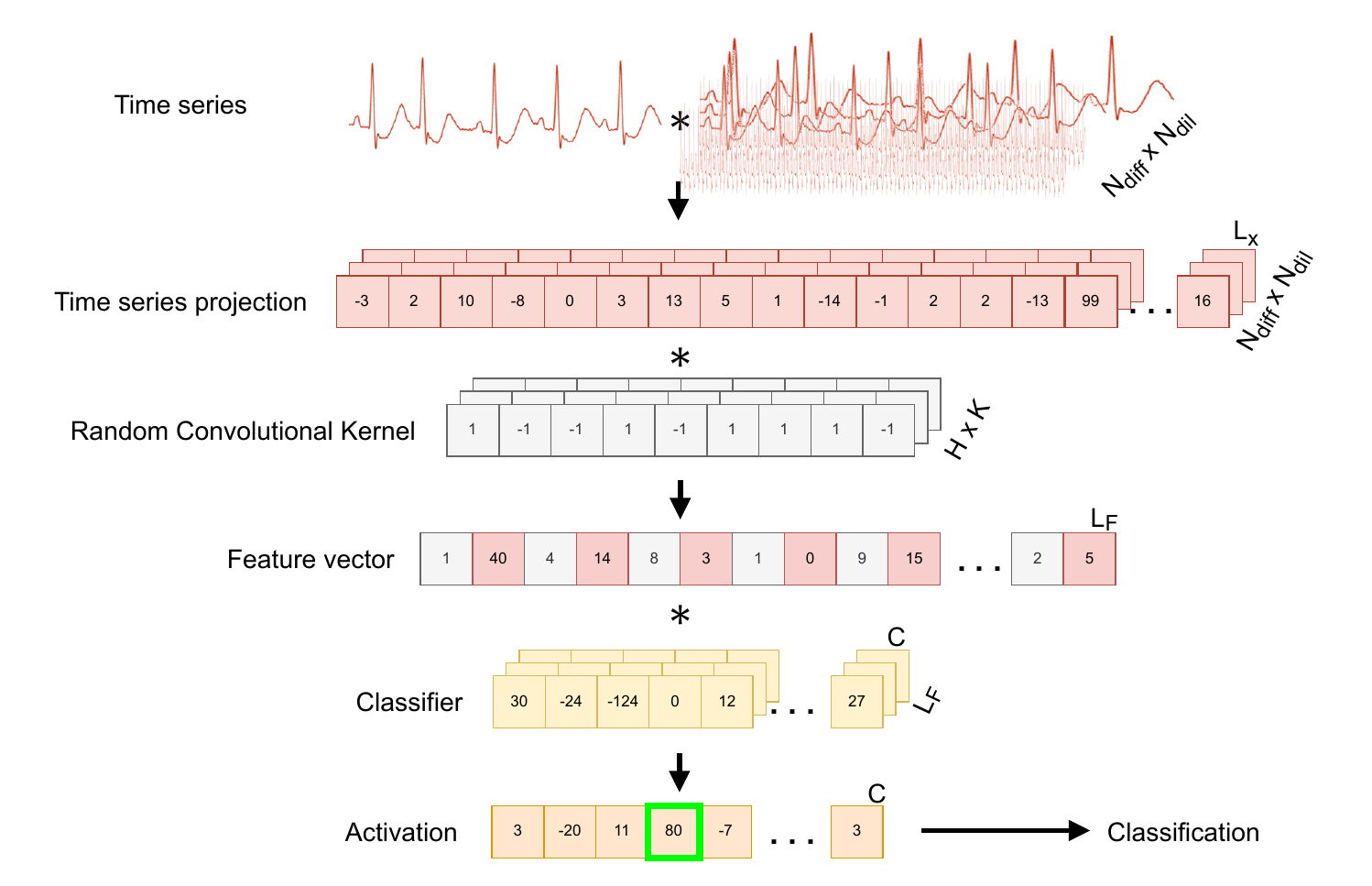}

\label{fig:overview}

\end{figure*}

\section{Background}
\label{sec:background}

Traditional \glspl{cnn} rely on convolutional kernels to extract meaningful features from the input data stream and thus capture temporal patterns.
The convolution operation consists of a sliding dot product between the kernel weights and the time series to be classified.
Such kernels are parametrized by their length, padding, stride, and dilation; a trade-off between accurate local representation and global understanding must be achieved, improving the classification accuracy while minimizing computational complexity.
To improve the latent representations, the weights and biases of the convolutional kernels are learned through a training process.

\glspl{rck} eliminate the need to train the parameters of the kernels that operate on the input features. 
\gls{rocket}~\cite{dempster2020rocket} relies on up to \rev{$10^{4}$} kernels for which the weights and biases, length, padding, and dilation are randomized, whereas the stride is fixed to 1.
The so-produced features are then passed through a linear classifier whose weights and biases are trained through \gls{sgd}.

\gls{hydra}~\cite{dempster2023hydra} makes use of \glspl{rck} to expand conventional dictionary techniques.
In \gls{hydra}, the transformation applied to the input stream is represented by the convolution with \glspl{rck}, with multiple kernels per group, one group for each value of the dilation parameter, effectively modeling multiple receptive field sizes.
Therefore, each group represents a dictionary, within which each kernel represents a known pattern.
At each timestep, the matches between the input and the kernels within each group are counted, and the resulting pattern occurrences are normalized and used to train a classifier.
Although accurate, the exclusive usage of \texttt{fp32} weights for both the random and trainable kernels and the costly floating-point operations limit the TinyML potential of Hydra.
Moreover, \gls{hydra} draws the \gls{rck} weights from a standard uniform distribution $\mathcal{N}(0,1)$.
However, as shown by MiniRocket~\cite{dempster2021minirocket}, \gls{rck}-based algorithms can use integer values sampled from $\{-2, -1, 0, 1\}$ without compromising the classification accuracy.
The representation range for the generated features is sufficient for the trainable classifier to produce accurate results.

\section{Methods}
\label{sec:methods}

Figure~\ref{fig:overview} depicts the NanoHydra algorithm designed to accurately and efficiently classify time series data. 
\rev{With the goal of enabling} \gls{tsc} at the extreme edge, we optimize the \gls{hydra} algorithm considering its hardware-associated costs.
We discuss each TinyML-associated component of the methodology in the following subsections.
 
\subsection{Random Convolutional Kernels}
\label{ssc:rck}

Let $x[i]$ be an input time series with a length of $L_{\text{x}}$ samples.
The input signal is convolved with differentiated and dilated versions of itself in order to eliminate offset effects and to extract patterns across different scales, respectively. 
The number of such versions constitutes two hyperparameters of the model, $\Ndf$ and $\Ndl$, that influence the computational effort of the analysis performed. 
All time series projections are then convolved with randomized kernels of length $W=9$, organized into $H$ groups of $K$ kernels each. 
That is, $w\sb{[h,k]}[j]$, where $k \in [0,\cdots, K-1]$, $h \in [0,\cdots, H-1]$ and $j \in [0,\cdots, 8]$. 
For each input sample $i \in [0, \cdots, X_{\text{len}}-1]$, a convolutional output $y\sb{[h,k,d]}[i] = \sum\sb{j=0}^{W-1} x[i+j\dot d] w\sb{[h,k]}[j]$ is produced.

Each $[h,k,\ndl,\ndf]$ convolution output is concatenated into a feature vector $F$, of length 

\begin{equation}
    L_{F} = \lambda \cdot H \cdot \Ndl \cdot \Ndf
\end{equation}

The feature vector can be interpreted as a histogram, where each $[h,k,\ndl,\ndf]$ feature is accumulated over its respective convolutional output for each time step $i$. 
Namely, the kernels with the maximum response and those with minimal similarity to the input time series are counted.
We set $\lambda = 2$, representing the accumulations performed over the feature vector.
Specifically, we increment the kernel count with an extreme response (i.e., hard counting), as well as accumulate the response itself (i.e., soft counting) at each timestep.

Note that the value of the convolutional output for the input sample at index $i$ is independent of the outputs at any other index $j \neq i$.
Therefore, it is not necessary to store all convolutional outputs $y_{[k,h,d]}[i]$ for each time-index $i$ of the input window during the counting process, thus significantly reducing resource usage.  
Instead, each feature vector $F$ update requires an output $y$ generated from a sample $i$ only $k \in [0,K-1]$ outputs.

\rev{To demonstrate NanoHydra, we first reduce the sampling set of the \gls{rck} weights to $\{-1, 1\}$, empirically shown by~\cite{dempster2020rocket} to produce accurate \gls{tsc} systems.}
By reducing the range of the kernel weights, we \rev{enable the replacement of} floating-point multiplications with efficient integer operations. 
Thus, we can reduce the on-device storage and memory requirements for NanoHydra, as \texttt{uint8} variables are sufficient to represent the target values.

\subsection{Sparse Scaling}
\label{ssc:sparsescaling}

In the offline, server-side training phase, each feature vector's mean $\mu$ and standard deviation $\sigma$ are calculated, as obtained from the training set. 
During the inference stage, the feature vectors $F\sb{[h,d]}[k]$ are normalized, ensuring a proportional contribution to the counting process.
Furthermore, a non-linear rectified linear unit activation function is applied, ensuring that only positive contributions are accounted for in the counting process.
By linearizing the access to $F$ over a new index $i$, such that $F\sb{[h,d]}[k] \equiv F[i]$, the features become:

\begin{align}
\label{eq:fscaled}
F\sb{\text{scaled}}[i] = \begin{cases} \frac{F[i] - \mu[i]}{\sigma[i]}, & \mbox{if } F[i] > 0 \\ 0, & \mbox{otherwise} \end{cases}
\end{align}

In the scaling stage, \gls{hydra}~\cite{dempster2023hydra} models apply the square root operator to the feature vectors produced through counting methods.
Such an operation reduces the magnitude of the values employed in the classification process.
We aim to avoid the computational effort required to perform floating-point arithmetics in embedded devices, whereas simply performing on-board approximations can lead to unseen feature distributions and, thus, incorrect predictions from the offline trained classifier. 
We thus replace the square root operator with an arithmetic shift during both the server-side training stage and the on-device inference process.

To the same aim, we replace division by the standard deviation detailed in Equation~\ref{eq:fscaled} with an arithmetic shift with its closest power-of-two value.
The normalization step performed at the feature level cannot be avoided, given the need to produce uniform values for the classifier.
As the values are produced through different counting strategies (i.e., soft counting accumulates convolutional activations, whereas hard counting accumulates occurrences), an arithmetic shift normalization improves the model convergence and training stability.
Instead of performing the normalization step after the mean subtraction, we propose directly scaling the convolutional outputs, discarding the least significant bits without compromising the accuracy.
This ensures that 16-bit signed integers are sufficient for the accumulation vectors without overflows, given the 16-bit signed integers operands of the convolution and considering a 32-bit buffer employed to temporarily store the result.
Furthermore, by replacing the division with the power-of-two arithmetic shift, each element $\sigma[i]$ can be stored as an 8-bit signed integer, saving 24 bits per element.

\subsection{Classification}
\label{ssc:classification_method}

To produce categorical class labels for the input time series, the features produced in Section \ref{ssc:sparsescaling} are fed to a linear classifier, followed by determining the class $c$ with the highest activation.
Namely:

\begin{equation}
    \text{classification} = \text{argmax}\sb{c} (B \cdot F\sb{\text{scaled}}),
\end{equation}

where $F\sb{\text{scaled}}$ is the scaled feature vector, and $B \in \mathcal{M}\sb{C \times L_{\sb{F}}}$ the dense matrix of trained weights for the logistic regression. 

The weights of the \gls{rck} are binary throughout the server-side training and on-device inference processes.
However, the trained classifier, which operates on the outputs of the \glspl{rck}, is frozen at deployment time and quantized to \texttt{int8} precision through post-training quantization.
Shown to preserve accuracy in classification tasks~\cite{mei2024train}, the integerization of the weights enables a $4\times$ reduction of the storage cost.
We opt to employ a 32-bit signed integer as the accumulation buffer for the matrix-vector multiplication, thus avoiding potential overflows caused by the \gls{mac} operations performed over a feature vector with up to thousands of elements~\cite{dempster2020rocket,dempster2021minirocket}.
As only one such accumulator per class is needed, the contribution to the total memory cost is minimal.

\subsection{Implementation}
\label{sec:nanohydra}

\begin{figure}[t]
\centering
\includegraphics[width=0.49\textwidth]{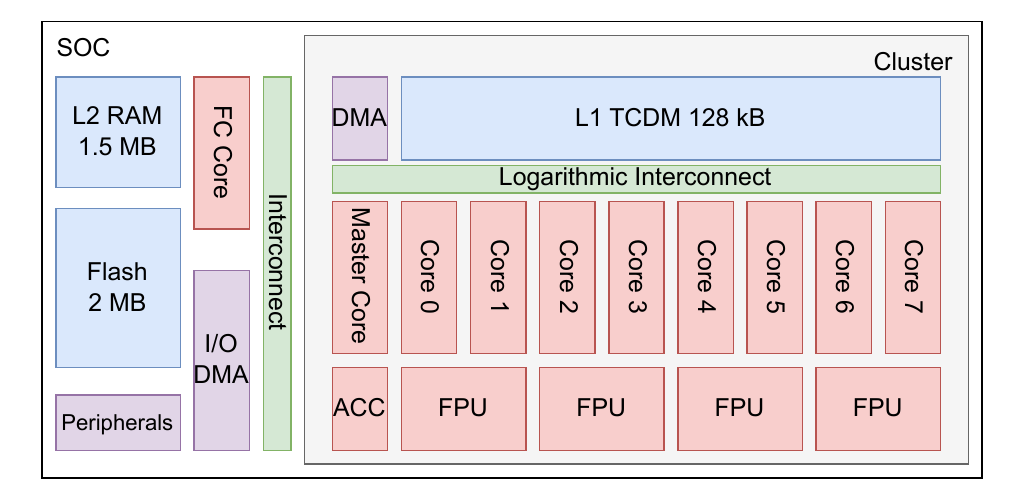}

\caption{\textbf{GAP9 MCU.} The block diagram depicts the computational resources and the memory levels available on the extreme edge device.}
\label{fig:gap9}
\end{figure}

We demonstrate the NanoHydra approach on the GAP9~\footnote{\url{https://greenwaves-technologies.com/gap9_processor/}} \gls{mcu} developed by Greenwaves Technologies, shown in Figure~\ref{fig:gap9}.
Two processing components comprise the \gls{mcu}: a fabric controller that handles the control and the communication processes and a 9-core cluster that can efficiently execute \glspl{mac}.
The cores share access to an \qty{128}{\kilo \byte} \gls{tcdm} memory, as well as four \glspl{fpu} that can perform single precision operations and a \gls{cnn} accelerator.
Additional on-chip memory (i.e., \qty{1.5}{\mega \byte} RAM and \qty{2}{\mega \byte} flash) and external memory interfaces are integrated into the GAP9 device for the deployment of \glspl{dnn} or the storage of on-site samples.

NanoHydra exploits the hardware features of the GAP9 \gls{mcu}.
Namely, we parallelize the \glspl{mac} performed by the convolutional layers, the linear classifier, and the feature vector computation.
Furthermore, we use the \gls{simd} instructions developed for the RISC-V cores on GAP9, as the entire inference process relies on quantized operators.
We can thus simultaneously perform four dot product operations within the linear layer with \texttt{int8} weights, as well as for the convolutional features generated by the \glspl{rck}.

\section{Results}
\label{sec:result}

\rev{As in \gls{rocket}~\cite{dempster2020rocket} and \gls{hydra}~\cite{dempster2023hydra} baselines, we} evaluate our NanoHydra methodology on the \rev{UCR Archive~\cite{hoanganh2018ucr}. 
We then focus on the ECG5000 dataset~\cite{hoanganh2018ucr} for further analyses.}
\rev{The former is a benchmark containing over 100 datasets, whereas the latter} consists of 5000 \gls{ecg} 1-second long data, each sampled at \qty{140}{\hertz}.
We employ only 500 \rev{ECG} samples in the training process.
The remaining 4500 represent the test set and are only seen at evaluation time, thus verifying NanoHydra's ability to generalize to unseen data.
The samples belong to five classes: Normal (N), R-on-T Premature Ventricular Contraction (R-on-T PVC), Premature Ventricular Contraction (PVC), Ectopic Beat (EB), and Unclassified Beat (UB).
We report the best accuracy measured over 20 runs for each model hyperparameter pair $\{\Ndl,\Ndf\}$.

As described in Section~\ref{sec:nanohydra}, we perform 4-way vectorization for the 8-bit \gls{rck} weights and the 8-bit logistic regression weights and parallelize the operations over the 8-core cluster.
We algorithmically validate our implementation using the GVSOC event-driven simulator and \rev{then} deploy our NanoHydra on the GAP9 \gls{mcu}.

\subsection{Classification Performance}
\label{ssc:classification}

\begin{table}[b]

\small

\caption{\revs{Average accuracy [\%] and standard deviation [\%] on the UCR Archive, as well as the average training time per dataset. For NanoHydra we report the best accuracy over 20 runs for each dataset. For related works, we show the results reported by~\cite{schafer2023weasel2}. Dictionary methods are marked with D and kernel-based methods are marked with K.}}
\label{tab:ucr}
% \vspace{0.3cm}

\centering

\begin{tabular}{@{}lrrr@{}}

% Dictionary (D): BOSS, cBOSS, WEASEL, TDE; 
% Hybrid (H): HiveCote 2.0, HiveCote 1.0, TS_CHIEF,
% Deep-Learning (DL): InceptionTime;
% Shapelets (S): R-DST, MrSQM_SFA_k5; 
% Kernel (K): Arsenal, MiniRocket, MultiRocket, Rocket, Hydra

% Topology         & Average [\%] & Std. dev. [\%] & \# datasets & Method \\
% BOSS             & 81.17 & 0.17 & 114 & D \\
% cBOSS            & 81.35 & 0.17 & 114 & D\\
% WEASEL           & 82.98 & 0.16 & 114 & D \\
% WEASEL2.0        & 86.30 & 0.14 & 114 & D \\
% TDE              & 83.99 & 0.15 & 114 & D\\ 
% % MrSQM\_SFA\_5k     & 85.00 & 0.14 & 114 & S/D\\
% Arsenal          & 85.41 & 0.16 & 112 & K \\
% % InceptionTime    & 87.15 & 0.13 & 108 & DL \\
% MiniRocket       & 86.09 & 0.14 & 114 & K \\
% Rocket           & 85.59 & 0.16 & 114 & K \\

% % R\_DST           & 86.32 & 0.14 & 114 & S \\
% % HC1              & 87.82 & 0.12 & 108 & H \\
% % TS-CHIEF         & 87.57 & 0.13 & 108 & H \\
% MultiRocket      & 86.84 & 0.14 & 113 & K \\
% HYDRA            & 85.31 & 0.15 & 114 & K/D \\
% % HC2              & 87.53 & 0.13 & 114 & H \\
% \textbf{NanoHydra (ours)} &\textbf{ 83.81} &\textbf{ 0.16} &\textbf{ 112} & \textbf{K/D} \\

\toprule
Topology         & Accuracy [\%] & Train time [s] & Method \\
\midrule
BOSS~\cite{schafer2015boss}             & 81.17 $\pm$ 0.17 & 1628.17 & D \\
WEASEL~\cite{schafer2017weasel}           & 82.98 $\pm$ 0.16 & 29.46 & D \\
cBOSS~\cite{middlehurst2019cboss}            & 81.35 $\pm$ 0.17 & 506.16 & D\\
Rocket~\cite{dempster2020rocket}           & 85.59 $\pm$ 0.16 & 67.41 & K \\
TDE~\cite{middlehurst2021tde}              & 83.99 $\pm$ 0.15 & 4261.99 &  D\\ 
Arsenal~\cite{middlehurst2021arsenal}          & 85.41 $\pm$ 0.16 & - & K \\
MiniRocket~\cite{dempster2021minirocket}       & 86.09 $\pm$ 0.14 & 10.65 & K \\
MultiRocket~\cite{tan2022multirocket}      & 86.84 $\pm$ 0.14 & 23.55 & K \\
HYDRA~\cite{dempster2023hydra}            & 85.31 $\pm$ 0.15 & 9.34 & K/D \\
WEASEL2.0~\cite{schafer2023weasel2}        & 86.30 $\pm$ 0.14 & 53.95 & D \\

\textbf{NanoHydra (ours)} &\textbf{83.81 $\pm$ 0.16} & \textbf{6.87} & \textbf{K/D}\\ % Q16
% \textbf{NanoHydra (ours)} &\textbf{82.33 $\pm$ 0.16} & \textbf{K/D} \\ % Q8

\bottomrule

\end{tabular}

\end{table}

First, we present the accuracy of \revs{an \texttt{int16} NanoHydra} on the datasets of the UCR Archive~\cite{hoanganh2018ucr} and compare it with other dictionary-based, kernel-based, and hybrid methodologies, shown in Table~\ref{tab:ucr}.
Despite being TinyML-optimized, our proposed model achieves competitive accuracy with respect to related works, outperforming \cite{schafer2015boss, schafer2017weasel, middlehurst2019cboss}.
\revs{While achieving a 3.03\% lower accuracy than the top-performing MultiRocket~\cite{tan2022multirocket} topology, our lightweight models are the fastest to train, with an average of \qty{6.87}{\second} per UCR Archive dataset, compared to the \qty{23.55}{\second} required for MultiRocket~\cite{tan2022multirocket}, as reported by Sch{\"a}fer et al.~\cite{schafer2023weasel2}.}

% \begin{table*}[!t]
% \renewcommand{\arraystretch}{1.3}
% \caption{Performance comparison of NanoHydra with other works on the ECG5000 Dataset.}
% \label{table:ours_vs_others}
% \centering
% \begin{tabular}{lrrrrrrr}
% \firsthline
% Algorithm            & \multicolumn{2}{c}{E-NanoHydra} & \multicolumn{2}{c}{A-NanoHydra} & 1-D TCN \cite{Zemouri2023} & ECG-TCN \cite{Ingolfsson2021} & NAS-TCN \cite{Burrello2023} \\
% \hline
% Memory  [kB]         & \multicolumn{2}{c}{3.82} & \multicolumn{2}{c}{12.57} & 52.43 & 26.63 & 15.2 \\
% Training Time [s]    & \multicolumn{2}{c}{38} & \multicolumn{2}{c}{95} & N/A & N/A & $10^{4}$ \\
% Accuracy     [\%]    & \multicolumn{2}{c}{94.13} & \multicolumn{2}{c}{94.47} & 95.1 & 93.8 & 94.2 \\
% \cmidrule(lr){1-1} \cmidrule(lr){2-3} \cmidrule(lr){4-5} \cmidrule(lr){6-6} \cmidrule(lr){7-7} \cmidrule(lr){8-8}
% Clock Speed [MHz]    & 100   & 240   & 100   & 240   & 100   & 100  & 100  \\
% Inference Time [ms]       & 0.733 & 0.328 & 3.354 & 0.984 & 121.4 & 2.70 & 2.69 \\
% Inference Energy [$\mu$J]  & 9.98  & 7.32  & 45.9  & 36.5  & N/A   & 100  & 140  \\
% %Battery Life [years] & 4 & 4.3 & 2.45 & 2.7 & N/A & 2 & N/A \\
% \lasthline
% \end{tabular}
% \end{table*}

\begin{table*}[t!]
\normalsize

\caption{NanoHydra performance on ECG5000 and comparison with state-of-the-art methods. NanoHydra is evaluated on GAP9 operating at \qty{100}{\mega \hertz} and \qty{650}{\milli \volt}.}
% \vspace{0.2cm}
\label{tab:classification}

\centering

\begin{tabular}{lrrrrr}

\toprule

Algorithm            & \multicolumn{1}{c}{E-NanoHydra} & \multicolumn{1}{c}{A-NanoHydra} & 1-D TCN \cite{zemouri2023embedded} & ECG-TCN \cite{ingolfsson2021ecgtcn} & NAS-TCN \cite{risso2023lightweight} \\

\midrule

Accuracy [\%]             & 94.13 & 94.47 & 95.1  & 93.8  & 94.2 \\
Memory [kB]               &  3.82 & 12.57 & 52.43 & 26.63 & 15.2 \\
Training time [s]         &  38   & 95    & N/A   & N/A   & $10^{4}$ \\
Inference time [ms]       & 0.77  & 2.23  & 121.4 & 2.70 & 2.69 \\
Inference energy [$\mu$J] & 10.10  & 31.48  & N/A   & 100  & 140 \\

\bottomrule

\end{tabular}

\end{table*}

\rev{
We then evaluate our NanoHydra proposal on the ECG5000 dataset, focusing on TinyML applications and constraints.
}
We consider \revs{two \texttt{int8} variants} of the topology, defined by the dilation values $\Ndl$ and the number of differentiations $\Ndf$.
We define Accurate-NanoHydra (A-NanoHydra) with $\{\Ndl = 5, \Ndf = 2\}$ and Efficient-NanoHydra (E-NanoHydra) with $\{\Ndl = 3, \Ndf = 1\}$.
The results are shown in Table \ref{tab:classification}.
A-NanoHydra achieves an accuracy of 94.47\% on the ECG5000 dataset, using only 500 training samples.
% dataset Hydra	Hydra+MiniRocket Hydra+MultiRocket Hydra+Rocket
% ECG5000 0.947133 0.947652	0.947348 0.948511
For comparison, \gls{hydra} baseline~\cite{dempster2023hydra} reports 94.85\% classification accuracy on the same dataset at a higher computational cost, \rev{whereas MiniRocket~\cite{dempster2021minirocket} reports an accuracy of 94.58\%.} 
We furthermore reduce the hardware-associated costs when employing E-NanoHydra, with a classification accuracy of 94.13\%.
Namely, only \qty{3.82}{\kilo \byte} of memory are required by the efficient topology of our method, $3.3\times$ lower than the peak memory consumption of A-NanoHydra, as we generate fewer time series projections which are further convolved with the \glspl{rck}.
Moreover, the training time reduces from \qty{95}{\second} to only \qty{38}{\second} by shrinking the model for a training set totaling \qty{500}{\second} of labeled time series.

We compare our NanoHydra topologies with state-of-the-art methods deployed on extreme edge devices.
For a fair comparison, we consider an operating point of \qty{100}{\mega \hertz} for each clock domain within the GAP9 platform. 
As shown in Table \ref{tab:classification}, the accuracy achieved by both NanoHydra model variants is comparable with the related work.
The advantages of our proposed methodology become evident when analyzing the on-device performance.
Notably, our memory requirements are $4 \times$ smaller than those of the automatically designed NAS-TCN~\cite{risso2023lightweight}, showing the efficiency of pairing \gls{rck} kernels with a dictionary approach.
Moreover, E-NanoHydra is $3.49 \times$ faster than NAS-TCN~\cite{risso2023lightweight} in producing an accurate prediction, thanks to the vectorization of the dot product operations and the multi-core parallelization of the \glspl{mac}.
We additionally benefit from the energy efficiency of the ultra-low-power GAP9, requiring only \qty{10}{\mu \joule} per inference, a tenth of the inference cost for ECG-TCN~\cite{ingolfsson2021ecgtcn}.

% Original

% \begin{table}[!h]

% \normalsize

% \caption{The impact of $N_{dil}$ and $N_{dif}$ on NanoHydra parallelization \\ on GAP9 \gls{mcu} operating at \qty{100}{\mega \hertz}.}
% \label{tab:sensitivity}
% \centering
% \begin{tabular}{ll | rrrr}

% % \cmidrule(lr){1-2} \cmidrule(lr){3-6}
% $N_{\text{dil}}$ & & \multicolumn{2}{c}{3} & \multicolumn{2}{c}{5}    \\
% % \cmidrule(lr){1-2} \cmidrule(lr){3-4} \cmidrule(lr){5-6}
% $N_{\text{diff}}$               & & 1 & 2 & 1 & 2 \\
% \multirow{2}{*}{Latency [ms]} &
%            1 core    & 4.86  & 9.66  & 8.09  & 16.1  \\
%           &  8-cores    & 0.73  & 1.36  & 1.16  & 2.21  \\
% Speed-up [$\times$]      & & 6.63  & 7.08  & 7.01  & 7.29  \\

% \end{tabular}
% \end{table}

% Reproduced

\begin{table}[b]

\normalsize

\caption{The impact of $\Ndl$ and $\Ndf$ on NanoHydra parallelization \\ on GAP9 \gls{mcu} operating at \qty{100}{\mega \hertz}.}
% \vspace{0.3cm}
\label{tab:sensitivity}
\centering
\begin{tabular}{ll | rrrr}

\toprule

% \cmidrule(lr){1-2} \cmidrule(lr){3-6}
$N_{\text{dil}}$ & & \multicolumn{2}{c}{3} & \multicolumn{2}{c}{5}    \\
% \cmidrule(lr){1-2} \cmidrule(lr){3-4} \cmidrule(lr){5-6}
$N_{\text{diff}}$               & & 1 & 2 & 1 & 2 \\

\hline
\midrule

\multirow{2}{*}{Latency [ms]} &
             1 core    & 4.99 & 9.96 & 8.32 & 16.60 \\
          &  8 cores &  0.77 & 1.43 & 1.21 & 2.23 \\
Speed-up [$\times$] &&  6.48 & 6.97 & 6.88 & 7.44  \\

\bottomrule

\end{tabular}
\end{table}

\begin{table}[b]
\footnotesize 

\caption{E-NanoHydra and A-NanoHydra inference cost per sample on the GAP9 MCU, considering four operating modes: ~\acrshort{ulpm} (i.e., $f = \qty{100}{\mega \hertz}$, $V_{\text{core}} = \qty{650}{\milli \volt}$), ~\acrshort{lpm} (i.e., $f = \qty{240}{\mega \hertz}$, $V_{\text{core}} = \qty{650}{\milli \volt}$) and~\acrshort{hpm} (i.e., $f = \qty{370}{\mega \hertz}$, $V_{\text{core}} = \qty{800}{\milli \volt}$).}
\label{tab:power}
% \vspace{0.3cm}

\centering

\begin{tabular}{lrrrrrr}

\toprule
Topology         & \multicolumn{3}{c}{E-NanoHydra} & \multicolumn{3}{c}{A-NanoHydra} \\
Operating mode            & ULPM & LPM & HPM & ULPM & LPM & HPM \\ 

\hline
\midrule
 
Power [mW]       & 13.09 & 23.02 & 51.24 & 13.69 & 25.63 & 60.11 \\

Latency [ms]     & 0.77 & 0.33 & 0.22 & 2.23 & 0.97 & 0.64 \\ 

Energy [$\mu$J]  & 10.10 & 7.69 & 11.50 & 31.48 & 24.86 & 38.26 \\

\bottomrule

\end{tabular}

\end{table}

\subsection{Sensitivity Analysis}
\label{ssc:sensitivity}

In Table~\ref{tab:sensitivity} we compare E-NanoHydra and A-NanoHydra with their intermediate counterparts, varying the number of differentiations and dilations employed to generate time series projections.
Once \rev{the} projections are fed to the convolutional kernels and, further down the pipeline, to the linear classifier, the number of cores employed in the computational process impacts the inference latency.
Note that all kernel weights are quantized to 8 bits, thus exploiting the \gls{simd} capabilities of GAP9~\gls{mcu}.
As the number of processing variants increases (i.e., from 3 for $\{\Ndl = 3, \Ndf = 1\}$ to 10 for $\{\Ndl = 5, \Ndf = 2\}$), so does the inference latency, from \qty{4.99}{\milli \second} to \qty{16.6}{\milli \second}. 
By parallelizing the computation over the multi-core cluster, the latency reduces to \qty{0.77}{\milli \second} and \qty{2.3}{\milli \second}. 
At the same time, the impact of the parallelization increases to $7.44\times$ as more intermediate features require more operations to produce a classification result.

We expand the on-device analysis presented in Section~\ref{ssc:classification} considering the operating conditions introduced by~\cite{cioflan2024ondevice}.
In Table~\ref{tab:power} we show the NanoHydra efficiency on the GAP9 \gls{mcu} for~\gls{ulpm}, i.e., $f = \qty{100}{\mega \hertz}$, $\Vcore = \qty{650}{\milli \volt}$, ~\gls{lpm}, i.e., $f = \qty{240}{\mega \hertz}$, $\Vcore = \qty{650}{\milli \volt}$ and~\gls{hpm}, i.e., $f = \qty{370}{\mega \hertz}$, $\Vcore = \qty{800}{\milli \volt}$.

The A-NanoHydra variants consume more power than any E-NanoHydra operating scenario since the increased number of operations requires the GAP9 \gls{mcu} to be active (i.e., not in deep sleep) for longer. 
Notably, the \gls{ulpm} requires more energy per inference than the \gls{lpm} since, despite operating at a lower average power, the latency of NanoHydra reduces the energy consumption.
Specifically, when analysing E-NanoHydra, we observe that $\Delta t_{\text{\gls{lpm}}} = 0.43\cdot\Delta t_{\text{\gls{ulpm}}}$ and $P_{\text{\gls{lpm}}} = 1.75 \cdot P_{\text{\gls{ulpm}}}$, and since $E = P_{\text{avg}}\Delta t$, we have that 

\begin{equation}
\label{eq:power_lat_scale}
\frac{E_{\text{\gls{lpm}}}}{E_{\text{\gls{ulpm}}}} = \frac{1.75 \cdot P_{\text{\gls{ulpm}}} \cdot 0.43\cdot\Delta t_{\text{\gls{ulpm}}} }{ P_{\text{\gls{ulpm}}} \cdot \Delta t_{\text{\gls{ulpm}}}} = 0.75,
\end{equation}
which demonstrates that \gls{lpm} offers an optimal latency-energy trade-off for both A-NanoHydra and E-NanoHydra.
When aiming to minimize the inference latency, at the expense of an increased energy cost, operating in \gls{hpm} represent the optimal solution given its \qty{370}{\mega \hertz} operating frequency.
This enables our \gls{tsc} system to offer accurate A-NanoHydra predictions in only \qty{0.64}{\milli \second}.

\subsection{Device lifetime}
\label{ssc:lifetime}

\begin{table}[b]
\small

\caption{Device lifetime estimates for two wearable system operating in~\gls{lpm} mode (i.e., $f = \qty{240}{\mega \hertz}$, $V_{\text{core}} = \qty{650}{\milli \volt}$): \gls{ulpws} and BioGAP~\cite{frey2023biogap9}. We abbreviate Efficient-NanoHydra to E-NH and Accurate-NanoHydra to A-NH}

\label{tab:batlife}

\centering

\begin{tabular}{lrrrr}

Wearable system        & \multicolumn{2}{c}{\gls{ulpws}} & \multicolumn{2}{c}{BioGAP~\cite{frey2023biogap9}} \\
Topology               & E-NH & A-NH & E-NH & A-NH\\ 

\hline
 
Device lifetime [years]       & 4.44 & 2.90 & 0.26 & 0.25\\

\end{tabular}

\end{table}

The device lifetime of an extreme edge platform integrated into a wearable system is a decisive metric in assessing the potential of such a system.
Regardless of the classification accuracy, a large power consumption could lead to \rev{unacceptably low battery lifetimes} and thus compromise user adoption.
In the following, we provide an analytical methodology to estimate the device lifetime of an \gls{tsc} system.
For active inference, we consider the power consumption associated with the computation $\Pinf$, the power consumption of the \gls{adc} integrated into the acquisition system $\Padc$, and the \gls{mcu} deep sleep power consumption $\Psl$.
Furthermore, we account for the inference computation time $\ti$ and the sample acquisition time $\ta$, as well as the battery capacity $\Bcap$ expressed in \qty{}{\milli \ampere \hour} and the voltage supply $\Vcore$. 

We assume that the acquisition system operates continuously ensuring that no time series data is lost, whereas the \gls{mcu} is active only during compute time, when sufficient input samples have been acquired and can be classified.
We can therefore estimate an average power consumption for the system given during an acquisition window:

\begin{equation}
    \Pavg = (\Pinf \ti + \Psl({\ta - \ti}))\frac{1}{\ta} + \Padc
\label{eq:pavg}
\end{equation}

We consider two types of wearable systems.
First, an ideal \gls{ulpws}, which integrates a GAP9 \gls{mcu} for the sample processing, operating in \gls{lpm} during compute time, as discussed in Section~\ref{ssc:sensitivity}, and consuming $\Psl = \qty{45}{\mu \watt}$ in deep sleep mode.
The proposed system additionally contains an ADS7042 12-bit off-the-shelf \gls{adc}~\cite{ADS7042tiadc} for the acquisition, the latter having a total power consumption of $\qty{1}{\mu \watt}$ with a sampling rate of \qty{1}{\kilo SPS}.
Second, we consider the medical-grade BioGAP Evaluation Board~\cite{frey2023biogap9}, a front-end acquisition board that features a GAP9 \gls{mcu} and an ADS1298 8-channel 24-bit \gls{adc}~\cite{ADS1298tiadc}.
The system consumes \qty{24}{\milli \watt} when performing computations on the edge node in \gls{lpm} and \qty{150}{\mu \watt} in sleep mode.

In Table~\ref{tab:batlife} we present the device lifetime estimates for the proposed wearable system, calculated using Equation~\ref{eq:pavg}.
We consider our proposed A-NanoHydra and E-NanoHydra topologies, deployed on the target platforms supplied by a CR2450 coin cell lithium battery, with $\Bcap = \qty{600}{\milli \ampere \hour}$ of charge capacity at $\Vbat = \qty{3}{\volt}$, suitable for TinyML systems~\cite{hou2023smg}.
Note that $\ta = \qty{1}{\second}$, as we classify samples from the ECG5000 dataset~\cite{hoanganh2018ucr}.

In an optimal scenario that neglects nonidealities such as DC-DC conversion losses or memory buffer consumptions, our A-NanoHydra network could offer accurate predictions for \qty{2.9}{years}.
At the expense of only 0.34\%, as shown in Section~\ref{ssc:classification}, the E-NanoHydra topology can operate continuously for almost four and a half years, reliably monitoring the state of a user while maximizing their comfort.
NanoHydra is also suitable for real-world systems, as our estimates show that deploying NanoHydra on the medical-grade biosignal acquisition and processing platform BioGAP~\cite{frey2023biogap9} enables the autonomous on-device prediction for over three months.
The use of compact and lightweight TinyML devices together with accurate and energy efficient algorithms paves the way to effective time series classification at the extreme edge.

\section{Conclusion}
\label{sec:conclusion}

We presented NanoHydra, a \gls{tsc} methodology aimed at extreme edge embedded systems.
Through the use of \glspl{rck} with integerized binary weights and \texttt{int8}-quantized classifier, we reduced the memory requirements of the system to under \qty{4}{\kilo \byte}.
We deployed our proposed algorithm on the ultra-low-power platform GAP9, exploiting its multi-core cluster and the \gls{simd} operators.
We enable heart condition monitoring in under \qty{1}{\milli \second} per \qty{1}{\second}-long \gls{ecg} sample, with an energy cost below \qty{10}{\mu \joule}.
We show that extreme edge devices employing NanoHydra can continuously perform time series classification between three months and over four years.  
NanoHydra achieves accuracy levels up to 94.47\%, enabling patient monitoring on intelligent sensor nodes.

\section*{Acknowledgment}
This work was partially supported by the Swiss State Secretariat for Education, Research, and Innovation (SERI) under the SwissChips initiative and the EU project dAIEDGE under grant No 101120726. We thank Jannis Sch\"onleber for the fruitful discussions.

\bibliography{main}
\bibliographystyle{IEEEtran}

\end{document}